\documentclass[12pt]{article}
\usepackage{epsfig}

\newcommand{\lsim} {\buildrel < \over {_\sim}}

\newcommand{\Odd}{${\cal O}$}
\newcommand{\pom}{{\cal P}}
\newcommand{\Pom}{{\cal P}}
\newcommand{\odd}{{\cal O}}
\newcommand{\etal} {{\em et al.}}
\newcommand{\NP}[1]{ Nucl.\ Phys.\ {\bf #1}}
\newcommand{\ZP}[1]{ Z.\ Phys.\ {\bf #1}}

\newcommand{\PL}[1]{ Phys.\ Lett.\ {\bf #1}}

\newcommand{\PRep}[1]{Phys.\ Rep.\ {\bf #1}}
\newcommand{\PR}[1]{Phys.\ Rev.\ {\bf #1}}
\newcommand{\PRL}[1]{ Phys.\ Rev.\ Lett.\ {\bf #1}}

\newcommand{\AmS}{{\protect\the\textfont2
  A\kern-.1667em\lower.5ex\hbox{M}\kern-.125emS}}

\begin {document}
\begin{flushright}
{\small
SLAC--PUB--8623\\
September 2000\\}
\end{flushright}

\bigskip
\begin{center}
{{\bf\LARGE   
Jet asymmetry in high energy diffractive production}
\footnote{Research partially supported by the Department of 
                Energy under contract
                DE--AC03--76SF00515, the European network Quantum Field Theory
                (1997-2000),
                and the Swedish Natural Science Research Council,
                contract F--PD 11264--301.}}

\bigskip
C. Merino\\
{\sl Departamento de F\'{\i}sica de Part\'{\i}culas,
        Universidade de Santiago de Compostela, \\
        Campus universitario s/n, 15706 Santiago de Compostela, Spain}\\
\medskip

S. J. Brodsky\\
{\sl Stanford Linear
        Accelerator Center\\
        Stanford University, Stanford, California 94309,USA}\\
 
\medskip

J. Rathsman\\
{\sl TH Division, CERN, \\
	CH-1211 Geneva 23, Switzerland}\\
 
\end{center}

\vfill

\begin{center}
{\bf\large   
Abstract }
\end{center}

We propose the asymmetry in the fractional
energy of
charm versus
anticharm jets produced in high energy diffractive photoproduction as
a sensitive test of the interference of the Odderon $(C = -)$ and Pomeron
$(C = +)$ exchange amplitudes in QCD. If measured at HERA,
this asymmetry
could provide the first experimental evidence of the Odderon.

\vfill

\begin{center} 
{\it Contribution to the Proceedings of the \\
International Euroconference in Quantum Chromodynamics: \\
15 Years of the QCD---Montpellier Conference (QCD 00) \\
    Montpellier, France \\
 6--12 July 2000 }\\
\end{center}

\vfill

\vfill\eject


\normalsize

\section{Introduction}

The existence of the Odderon, an odd charge-conjugation, zero
flavor-number exchange
contribution to high energy hadron scattering
amplitudes was already discussed many years ago~\cite{Gribov}.
In Regge theory, the Odderon contribution is dual to a sum over
$C = P= -1$ gluonium states in the $t$-channel~\cite{lukaszuk_nicolescu}.
Also in quantum chromodynamics, the Odderon is a basic prediction following
simply from the existence of the color-singlet exchange
of three reggeized gluons in the
$t-$channel~\cite{kwiencinski_bartels}.
For reactions
which involve high momentum transfer, the deviation of the Regge
intercept of the Odderon trajectory from $\alpha_{\cal O}(t=0)=1$ can
in principle be computed~\cite{Lipatov2,Braun,Wosieck,Gauron2}
from perturbative QCD in
analogy to the methods used to compute the properties of the hard BFKL
Pomeron~\cite{BFKL}.

In the past, some tests were proposed to either verify or reject the odderon
hypothesis (e.g., to check~\cite{Contogouris} out
the Odderon hypothesis
in the range $0.05\lsim -t\lsim 1 Gev^{2}$,
by taking into account
a combination of differential
cross-sections for $\pi^{\pm}p\rightarrow\rho^{\pm}p$ and
$\pi^{-}p\rightarrow\rho^0n$, or of inclusive cross sections for
$\pi^{\pm}p\rightarrow\rho X$),
but no experimental
evidence
of the Odderon separate existence has ever been published. Now that
the recent
results from the
electron-proton collider experiments at
HERA~\cite{HERA} have brought renewed interest in
the nature and behavior of both the Pomeron~\cite{Brodsky1,CKMT1}
and the Odderon,
we propose~\cite{ourpaper} an experimental test well suited to HERA
kinematics which should
be able to disentangle the contributions of both the Pomeron and the
Odderon to  diffractive production of charmed jets.

\section{Odderon-Pomeron interference}

Consider the amplitude for diffractive
photoproduction of a charm quark anti-quark pair.
The leading diagram is given by single Pomeron exchange~(two reggeized
gluons), and the  next term in the Born expansion is given by the exchange
of one  Odderon~(three reggeized gluons). 
Both diffractive photoproduction and leptoproduction can be considered, 
although in the
following we will specialize to the case of photoproduction for
which the rate observed at HERA is much larger. Our results can easily be
generalized to non-zero $Q^2$.

We use the conventional kinematical variables, and we denote by
$z_{c(\bar{c})}$
the energy sharing of the $c\bar{c}$ pair
($z_{c}+z_{\bar{c}}=1$
in Born approximation at the parton level), and we take into account 
that the finite charm quark mass
restricts the range of $z$.
Moreover, $\xi$ is effectively the longitudinal momentum
fraction of the proton carried by the Pomeron/Odderon, and the proton mass
is neglected.

Regge theory, which is applicable in the kinematic region
$s_{\gamma p} \gg M_X^2 \gg M_Y^2$,
together with crossing symmetry, predicts the phases and analytic
form of high energy amplitudes (see, for example,
Refs.~\cite{collins} and~\cite{kaidalov}).
The amplitude for the diffractive process
$\gamma p \to c\bar{c} p^\prime$ with Pomeron (\Pom) or Odderon (\Odd)
exchange can be written as
\begin{equation}\label{eq:ampl}
{\cal M}^{\pom/\odd}(t,s_{\gamma p},M_X^2,z_c) \propto 
 g_{pp^\prime}^{\pom/\odd}(t)
\left(\frac{s_{\gamma p}}{M_X^2}\right)^{\alpha_{\pom/\odd}(t)-1}
\frac{\left(1+S_{\pom/\odd}e^{-i\pi\alpha_{\pom/\odd}(t)}\right)}
{\sin\pi\alpha_{\pom/\odd}(t)}
g_{\pom/\odd}^{\gamma c\bar{c}}(t,M_X^2,z_c)
\end{equation}
where $S_{\pom/\odd}$ is the signature (even (odd) signature
corresponds to an exchange which is (anti)symmetric under the interchange
$s\leftrightarrow~u$), which is $+(-)1$ for the Pomeron (Odderon).
In the Regge approach the upper vertex
$g_{\pom/\odd}^{\gamma c\bar{c}}(t,M_X^2,z_c)$
can be treated as a
local real coupling such that the phase is contained in the signature
factor.  In the same way the factor $g_{pp^\prime}^{\pom/\odd}(t)$
represents the lower vertex.

In general the Pomeron and Odderon exchange
amplitudes will  interfere, as illustrated in Fig.~1.
The contribution of the
interference term to the total  cross-section is  zero, but it
does contribute to charge-asymmetric rates. Thus we propose to study
photoproduction of $c$-$\bar{c}$ pairs and measure the asymmetry in
the energy fractions $z_c$ and $z_{\bar{c}}$.  More generally,
one can use other charge-asymmetric kinematic configurations, as well as
bottom or strange quarks.

\begin{figure}[htb]
\begin{center}
\epsfig{figure=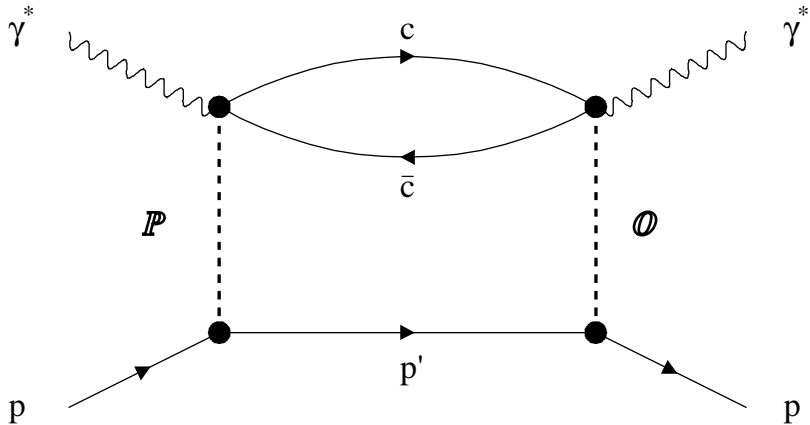}
\caption[*]{The interference between Pomeron (\Pom) or 
Odderon (\Odd)
exchange in the diffractive process $\gamma p \to c\bar{c} p^\prime$.}
\end{center}
\label{fig:interference}
\end{figure}

Given the amplitude~(\ref{eq:ampl}), the contribution to the cross-section from the
interference term depicted in Fig.~1 is proportional to
\begin{eqnarray} \label{eq:int}
\frac{d\sigma^{int}}{dtdM_X^2dz_c} 
&\propto& 
{\cal M}^{\pom}(t,s_{\gamma p},M_X^2,z_c)
\left\{{\cal M}^{\odd}(t,s_{\gamma p},M_X^2,z_c) \right\}^\dagger +h.c.
\nonumber \\ 
&=& 
g_{pp^\prime}^{\pom}(t)g_{pp^\prime}^{\odd}(t)
\left(\frac{s_{\gamma p}}{M_X^2}\right)^{\alpha_{\pom}(t)+\alpha_{\odd}(t)-2}
\frac{2\sin \left[ \frac{\pi}{2}\left(\alpha_{\odd}(t)-\alpha_{\pom}(t)\right)
\right]}
{\sin{ \frac{\pi\alpha_\pom(t)}{2}} \cos{ \frac{\pi\alpha_\odd(t)}{2}}
\nonumber}\\
[1ex]\nonumber\\ 
& & \times
g_{\pom}^{\gamma c\bar{c}}(t,M_X^2,z_c)
g_{\odd}^{\gamma c\bar{c}}(t,M_X^2,z_c) \ .
\end{eqnarray}
In the same way we can obtain the contributions to the cross-section from the
non-interfering terms for Pomeron and Odderon exchange.

The interference term can then be isolated by forming
the charge asymmetry,
\begin{equation} \label{eq:asym}
{\cal A}(t,M_X^2,z_c) = 
\frac{\displaystyle \frac{d\sigma}{dtdM_X^2dz_c}
                  - \frac{d\sigma}{dtdM_X^2dz_{\bar{c}}} }
{\displaystyle \frac{d\sigma}{dtdM_X^2dz_c}
             + \frac{d\sigma}{dtdM_X^2dz_{\bar{c}}} } \; .
\end{equation}
Thus the predicted asymmetry turns out to be
\begin{eqnarray}\label{eq:pred}
&&{\cal A}(t,M_X^2,z_c)  =  
{\displaystyle g_{pp^\prime}^{\pom}
g_{pp^\prime}^{\odd}
\left(\frac{s_{\gamma p}}{M_X^2}\right)^{\alpha_{\pom}+\alpha_{\odd} }}
{\displaystyle
\frac{2\sin \left[\frac{\pi}{2}\left(\alpha_{\odd}-\alpha_{\pom}\right)
 \right]}
{\sin  \frac{\pi\alpha_\pom}{2} \cos \frac{\pi\alpha_\odd}{2}}
\, g_{\pom}^{\gamma c\bar{c}}
g_{\odd}^{\gamma c\bar{c}}}\nonumber\\
& &\times
\left[\left(
g_{pp^\prime}^{\pom}
\left(
\frac{s_{\gamma p}}{M_X^2}
\right)^{\alpha_{\pom} }
g_{\pom}^{\gamma c\bar{c}}
/\sin\frac{\pi\alpha_\pom}{2}
\right)^2 
+
\left(g_{pp^\prime}^{\odd}
\left(\frac{s_{\gamma p}}{M_X^2}\right)^{\alpha_{\odd} }
g_{\odd}^{\gamma c\bar{c}}
/\cos\frac{\pi\alpha_\odd}{2}
\right)^2\right]^{-1}  \ .
\end{eqnarray}

The main functional dependence in the different
kinematic variables is expected to come from different factors in
the asymmetry. Thus,
the invariant mass $M_X$ dependence is mainly given by the power
behavior, $\left({s_{\gamma p}}/{M_X^2}\right)^{ \alpha_{\odd}(t)
-\alpha_{\pom}(t)}$, and it will thus  provide direct information about the
difference between $\alpha_{\odd}$ and $\alpha_{\pom}$. Another interesting
question which can be addressed from observations of the asymmetry is  the
difference in the $t$-dependence of $g_{pp^\prime}^{\odd}$ and
$g_{pp^\prime}^{\pom}$.

Let's note that in a perturbative calculation at tree-level
the interference would be zero
in the high-energy limit $s\gg|t|$ since the two- and three-gluon exchanges
are purely imaginary and real respectively. This should be compared with
the analogous QED process, $\gamma Z \to \ell^+\ell^- Z$, where the
interference of the one- and two-photon exchange amplitudes
can explain~\cite{Brodsky2} the observed lepton asymmetries, energy
dependence, and nuclear target dependence of the experimental
data~\cite{Ting} for large angles.

Also it is important to mention that secondary Regge
trajectories with the same quantum numbers as the Odderon
contribute, in principle, to the asymmetry at the current energies.
In fact, the $\omega$ contribution seems to be present in the fits to some of
the available experimental data. However, due to the intercept values of these
secondary Regge trajectories, their contribution will become negligible at
higher energies.

The ratio of the Odderon and Pomeron couplings to the proton,
$g_{pp^\prime}^{\odd}/ g_{pp^\prime}^{\pom}$,
is limited by data on
the difference of the elastic proton-proton and proton-antiproton
cross-sections at large energy $s$.
Following~\cite{kilian_nachtmann} we use the estimated limit on
the difference between the ratios of the real and imaginary part
of the proton-proton and proton-antiproton forward amplitudes,
\begin{equation}
\left|\Delta \rho(s)\right|=
\left|\frac{\Re \{ {\cal M}^{pp}(s,t=0)\} }
{\Im \{ {\cal M}^{pp}(s,t=0)\} } -
\frac{\Re \{ {\cal M}^{p\bar{p}}(s,t=0)\} }
{\Im \{ {\cal M}^{p\bar{p}}(s,t=0)\} }
\right| \leq 0.05 
\end{equation} 
for $s\sim10^4$ GeV$^2$ to get a limit on the ratio of the
Odderon and Pomeron couplings to the proton.
Using the amplitude corresponding to Eq.~(\ref{eq:ampl}) for proton-proton
and proton-antiproton scattering we get for $t=0$,
\begin{equation}
\Delta \rho(s) = 
2\frac{\Re \{ {\cal M}^{\odd}(s)\} }
{\Im \{ {\cal M}^{\pom}(s)\} + \Im \{ {\cal M}^{\odd}(s)\}}
\simeq -2\left(\frac{g_{pp^\prime}^{\odd}}{ g_{pp^\prime}^{\pom}} 
\right)^2
\left(\frac{s}{s_0}\right)^{\alpha_{\odd} -\alpha_{\pom}}
\tan{\frac{\pi\alpha_\odd}{2}} ,
\end{equation}
where $s_0$ is a typical hadronic scale $\sim 1$ GeV$^2$
which replaces $M_X^2$ in Eq.~(\ref{eq:ampl}). In the last step we
also make the simplifying assumption that the contribution to the denominator
from the Odderon is numerically much smaller
than from the Pomeron and therefore can be neglected.
The maximally allowed Odderon coupling at t=0 is then given by,
\begin{equation}\label{eq:rholim}
\left|g_{pp^\prime}^{\odd}\right|_{\max}=
\left|g_{pp^\prime}^{\pom}\right| 
\sqrt{\frac{\Delta \rho_{\max}(s)}{2} \cot \frac{\pi\alpha_{\odd}}{2}
\left(\frac{s}{s_0}\right)^{\alpha_{\pom} -\alpha_{\odd}}} .
\end{equation}
Strictly speaking this limit applies for the soft Odderon and Pomeron and
are therefore not directly applicable to charm photoproduction which
is a harder process, {\it i.e.} with larger energy dependence.
Even so we will use this limit to get an estimate of the maximal 
Odderon coupling to the proton.

\begin{figure}[htb]
\begin{center}
{\epsfig{figure=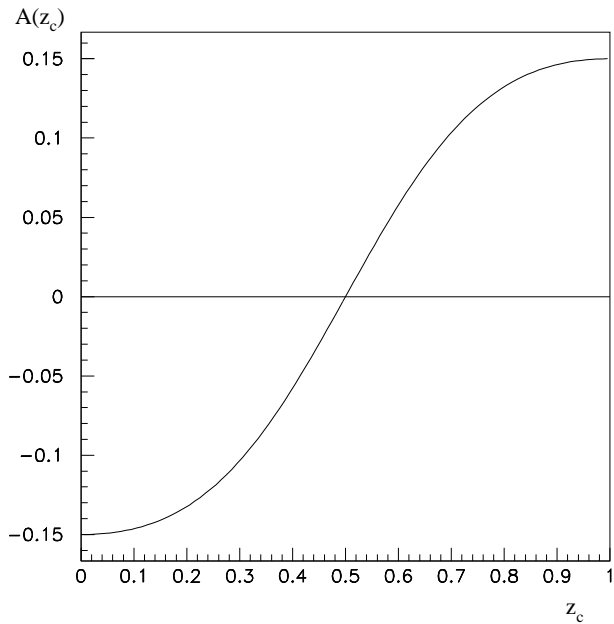,width=8.5cm}}
\caption{The asymmetry in fractional energy $z_c$ of 
charm versus anticharm jets predicted by our model using the
Donnachie-Landshoff Pomeron for $ \alpha_{\pom} = 1.2$,
$\alpha_{\odd} = 0.95$ and ${s_{\gamma p}}/{M_X^2}=100$.}
\end{center}
\label{fig:asym}
\end{figure}

The amplitudes for the asymmetry 
can be calculated using the Donnachie-Lands\-hoff~\cite{donnachie_landshoff}
model for the Pomeron and a similar ansatz for the
Odderon~\cite{kilian_nachtmann}.
The coupling of the Pomeron/Odderon to a quark is then given by
$\kappa^{\gamma c\bar{c}}_{\pom/\odd}\gamma^\rho$,
{\it i.e.} assuming a helicity preserving local interaction.
In the same way the Pomeron/Odderon couples to the proton with
$3\kappa_{pp^\prime}^{\pom/\odd}F_1(t)\gamma^\sigma$ if we only include
the Dirac form-factor $F_1(t)$.
The amplitudes
for the asymmetry can then be obtained by replacing
$g_{pp^\prime}^{\pom/\odd}(t)g_{\pom/\odd}^{\gamma c\bar{c}}(t,M_X^2,z_c)$
in Eq.~(\ref{eq:ampl}) by,
\begin{eqnarray*}
g_{pp^\prime}^{\pom/\odd}(t) g_{\pom/\odd}^{\gamma c\bar{c}}(t,M_X^2,z_c)
&=&
3\kappa_{pp^\prime}^{\pom/\odd}F_1(t)\bar{u}(p-\ell)\gamma^{\sigma}u(p)
\left(g^{\rho\sigma} -
\frac{\ell^\rho q^\sigma  + \ell^\sigma q^\rho }{\ell q}\right)
\kappa^{\gamma c\bar{c}}_{\pom/\odd}\epsilon^{\mu}(q)
\nonumber \\ &\times&
\bar{u}(p_c)
\left\{
\gamma^\mu \frac{\not\ell \,- \not p_{\bar{c}}+m_c}{(1-z)M_X^2}\gamma^\rho
-S_{\pom/\odd}
\gamma^\rho \frac{\not p_{c} \,- \not \ell+m_c}{zM_X^2}\gamma^\mu
\right\}v(p_{\bar{c}})
\end{eqnarray*}
where $\ell=\xi p$ is the Pomeron/Odderon momentum and
$g^{\rho\sigma} -\frac{\ell^\rho q^\sigma  + \ell^\sigma q^\rho }{\ell q}$
stems from the Pomeron/Odderon ``propagator". Note the signature
which is inserted for the crossed diagram to model the charge
conjugation property of the Pomeron. The Pomeron amplitude written
this way is not gauge invariant and therefore we use radiation
gauge also for the photon, {\it i.e.} the polarization sum is obtained using
$g^{\mu\nu}-\frac{q^\mu p^\nu + q^\nu p^\mu}{pq}$.
The leading terms in a $t/M_X^2$ expansion of the
squared amplitudes for the Pomeron and Odderon exchange as well
as the interference are then given by,
\begin{eqnarray}
\left(\frac{g_{pp^\prime}^{\pom}g_{\pom}^{\gamma c\bar{c}}}
{\kappa_{pp^\prime}^{\pom}\kappa_\pom^{\gamma c\bar{c}}}\right)^2
& \propto &
\frac{z_c^2+z_{\bar{c}}^2}{z_cz_{\bar{c}}}\frac{(1-\xi)}{\xi^2}
\nonumber \\[1ex]
\left(\frac{g_{pp^\prime}^{\odd}g_{\odd}^{\gamma c\bar{c}}}
{\kappa_{pp^\prime}^{\odd}\kappa_\odd^{\gamma c\bar{c}}}\right)^2
& \propto &
\frac{z_c^2+z_{\bar{c}}^2}{z_cz_{\bar{c}}}\frac{(1-\xi)}{\xi^2}
\nonumber \\[1ex]
\frac{g_{pp^\prime}^{\pom}g_{pp^\prime}^{\odd}g_{\pom}^{\gamma c\bar{c}}
g_{\odd}^{\gamma c\bar{c}}}
{\kappa_{pp^\prime}^{\pom}\kappa_{pp^\prime}^{\odd}
\kappa_\pom^{\gamma c\bar{c}}\kappa_\odd^{\gamma c\bar{c}}}
& \propto &
\frac{z_c-z_{\bar{c}}}{z_cz_{\bar{c}}}\frac{(1-\xi)}{\xi^2} ,
\end{eqnarray}
with corrections that are of order $t/M_X^2$ and therefore can be safely
neglected. The ratio
between the interference term and the Pomeron exchange is thus given by 
\begin{equation}\label{eq:upper}
\frac{g_{pp^\prime}^{\odd}g_{\odd}^{\gamma c\bar{c}}}
     {g_{pp^\prime}^{\pom}g_{\pom}^{\gamma c\bar{c}}}
=\frac{\kappa_{pp^\prime}^{\odd}\kappa_{\odd}^{\gamma c\bar{c}}}
{\kappa_{pp^\prime}^{\pom}\kappa_\pom^{\gamma c\bar{c}}}
\frac{z_c-z_{\bar{c}}}{z_c^2+z_{\bar{c}}^2}
=\frac{\kappa_{pp^\prime}^{\odd}\kappa_{\odd}^{\gamma c\bar{c}}}
{\kappa_{pp^\prime}^{\pom}\kappa_\pom^{\gamma c\bar{c}}}
\frac{2z_c-1}{z_c^2+(1-z_c)^2} \ .
\end{equation}
Inserting this into the expression of the asymmetry 
and making the simplifying assumption that the
Odderon contribution can be dropped in the denominator gives
\begin{equation}\label{eq:simp}
{\cal A}(t,M_X^2,z_c) \simeq 2
\frac{\kappa_{pp^\prime}^{\odd}\kappa_{\odd}^{\gamma c\bar{c}}}
{\kappa_{pp^\prime}^{\pom}\kappa_\pom^{\gamma c\bar{c}}}
\sin \left[\frac{\pi\left(\alpha_{\odd}-\alpha_{\pom}\right)}{2}\right]
\left(\frac{s_{\gamma p}}{M_X^2}\right)^{\alpha_{\odd}-\alpha_{\pom}}
\frac{\sin{ \frac{\pi\alpha_\pom}{2}}}{\cos{ \frac{\pi\alpha_\odd}{2}}}
\frac{2z_c-1}{z_c^2+(1-z_c)^2} \;.
\end{equation}
To obtain a numerical estimate of the asymmetry, we
shall assume that $t\simeq 0$ and use
$ \alpha_{\pom}^{hard} = 1.2$  and $\alpha_{\odd} = 0.95$
\cite{Wosieck} for the Pomeron and Odderon intercepts,
even though a recent paper by J.~Bartels, L.N.~Lipatov, and 
G.P.~Vacca~\cite{vacca}
has presented an Odderon solution in perturbative QCD with precise 
symmetry properties and intercept one.
In addition we will also assume
${\kappa_{\odd}^{\gamma c\bar{c}} }/{\kappa_\pom^{\gamma c\bar{c}} }
\sim \sqrt{C_F\alpha_s(m_c^2)} \simeq 0.6$, motivated by the Davies, Bethe,
and Maximon calculation~\cite{Bethe},
and use the maximal Odderon-proton coupling,
$\kappa_{pp^\prime}^{\odd}/\kappa_{pp^\prime}^{\pom} =
g_{pp^\prime}^{\odd}/g_{pp^\prime}^{\pom}=0.1$,
which follows from Eq.~(\ref{eq:rholim}) for
$ \alpha_{\pom}^{soft} = 1.08$, $s=10^4$ GeV$^2$,
$s_0=1$ GeV$^2$ and $\Delta \rho_{\max}(s)=0.05$.
Inserting the numerical values discussed above then gives
\begin{equation}
{\cal A} (t\simeq 0,M_X^2,z_c) \simeq\nonumber
0.45 \; \left(\frac{s_{\gamma p}}{M_X^2}\right)^{-0.25} \;
\frac{2z_c-1}{z_c^2+(1-z_c)^2}\; ,
\end{equation}
which for a typical value of $\frac{s_{\gamma p}}{M_X^2}=100$ becomes
a $\sim15$ \% asymmetry for large $z_c$ as illustrated in Fig.~\ref{fig:asym}.
Let's note that the asymmetry can also be integrated over $z_c$.

It should be emphasized that the magnitude of this estimate is quite
uncertain, since we are using an Odderon coupling to the proton which is
maximal for the soft Odderon in relation to the soft Pomeron, and
the hard Odderon and Pomeron could have a different ratio for the
coupling to the proton.

\section{Conclusions}
By observing the charge asymmetry of the quark/antiquark energy fraction
($z_c$)
in diffractive $c\bar{c}$ pair photoproduction or electroproduction,
the interference
between the Pomeron and the Odderon exchanges can be isolated and the
ratio to the sum of the Pomeron and the Odderon exchanges can be measured.
Using a  model with helicity conserving coupling for the Pomeron/Odderon
to quarks, the asymmetry is predicted to be proportional to
$(2z_c-1)/(z_c^2+(1-z_c)^2)$. The magnitude of the asymmetry is estimated
to be of order 15\%. However this estimate includes several unknowns and
is thus quite uncertain. This test could be performed
by current experiments at HERA, and possibly by COMPASS and STAR,
providing the first experimental evidence for the
existence of the Odderon.

\section{Acknowledgments}
It's a pleasure to thank M.A. Braun for his useful comments.


\begin{thebibliography}{9}

\bibitem{Gribov} V.N. Gribov, V.D. Mur, I.Yu. Kobzarev, L.B. Okun, and V.S.
Popov, \PL{B32} (1970) 129.

\bibitem{lukaszuk_nicolescu}
L.~Lukaszuk and B.~Nicolescu, Nuovo Cimento Letters {\bf 8} (1973) 405.

\bibitem{kwiencinski_bartels}
J. Kwiecinski and M. Praszalowicz, Phys. Lett. {\bf 94B} (1980) 413;
J. Bartels, Nucl. Phys. {\bf B175} (1980) 365.

\bibitem{Lipatov2}
P.~Gauron, L.N.~Lipatov, and B.~Nicolescu, \ZP{C63} (1994) 253.

\bibitem{Braun}
N.~Armesto and M.A.~Braun, \ZP{C75} (1997)  709.

\bibitem{Wosieck}
R.A.~Janik and J.~Wosieck, \PRL{82} (1999) 1092.

\bibitem{Gauron2}
M.A.~Braun, P.~Gauron, and B.~Nicolescu, hep-ph/9809567.

\bibitem{BFKL}
E.A.~Kuraev, L.N.~Lipatov, and V.S.~Fadin, Sov. Phys. JETP
{\bf44} (1976) 443; Sov. Phys. JETP {\bf 45} (1977) 199.\\
Y.Y.~Balitski and L.N.~Lipatov, Sov. J. Nucl. Phys. {\bf 28} (1978) 822.

\bibitem{Contogouris} A.P. Contogouris, L. Jenkovsky, E Martynov, and
B. Struminsky, \PL{B298} (1993) 432.

\bibitem{HERA}
H1 Collaboration, T.~Ahmed \etal, \NP{B429}, 477
(1994); \PL{B348} (1995) 681; C.~Adloff \etal, \ZP{C74} (1997)
221.\\
ZEUS Collaboration, M.~Derrick \etal, \PL{B315} (1993) 481;
J.~Breitweg \etal, \ZP{C75} (1997) 421; Eur. Phys. J.
{\bf C2} (1998) 237.

\bibitem{Brodsky1}
S.J.~Brodsky, L.~Frankfurt, J.F.~Gunion, A.H.~Mueller, and M.~Strikman,
\PR{D50} (1994) 3134.

\bibitem{CKMT1}
A.~Capella, A.B.~Kaidalov, C.~Merino, and J.~Tran Thanh Van,
\PL{B337} (1994) 358;\\
A.~Capella, A.B.~Kaidalov, C.~Merino, D.~Petermann, and J.~Tran Thanh Van,
\PR{D53} (1996) 2309.

\bibitem{ourpaper}
S.J. Brodsky, J. Rathsman and C. Merino, \PL{B461} (1999) 114.

\bibitem{collins}
P.D.B.~Collins, An introduction to Regge theory and high energy physics,
Cambridge University Press (1977).

\bibitem{kaidalov}
A.B.~Kaidalov, \PRep{50} (1979) 157.

\bibitem{Brodsky2}
S.J.~Brodsky and J.~Gillespie, \PR{173} (1968) 1011.

\bibitem{Ting}
S.C.C.~Ting, Proceedings of the International School of
Physics Ettore Majorana, Erice (Trapani), Sicily, July 1967.\\
J.G.~Ashbury \etal, \PL{B25} (1967) 565.

\bibitem{kilian_nachtmann}
W.~Kilian and O.~Nachtmann, Eur. Phys. J. {\bf C5} (1998) 317.

\bibitem{donnachie_landshoff}
A. Donnachie and P.V. Landshoff, Nucl. Phys. {\bf B244} (1984) 322;
ibid. {\bf B267} (1986) 690; Phys. Lett. {\bf B185} (1987) 403.

\bibitem{vacca} J. Bartels, L.N. Lipatov and G.P. Vacca, hep-ph/9912423.

\bibitem{Bethe}
H.A.~Bethe and L.C.~Maximon, Phys. Rev.~{\bf 93} (1954) 768;
H.~Davies, H.A.~Bethe, and L.C.~Maximon, Phys.~Rev.~{\bf 93} (1954) 788.

\end{thebibliography}
\end{document}